\documentclass{elsart1p}

\usepackage{graphicx}

\usepackage{amssymb}

\begin{document}

\vspace{-6cm}

\begin{frontmatter}



\title{The Qweak Experiment: a Search for New Physics at the TeV Scale
\thanksref{support}}
\thanks[support]{This work is supported in part by the US DoE, the US NSF,
NSERC (Canada), Jefferson Laboratory, and TRIUMF}


\author[addr1,addr2]{Willem T.H. van Oers} 
\author{(for the Qweak Collaboration)}

\address[addr1]{University of Manitoba, Winnipeg, MB, Canada R3T 2N2}
\address[addr2]{TRIUMF, 4004 Wesbrook Mall, Vancouver, BC, Canada V6T 2A3}

\begin{abstract}

A new precision measurement of the parity violating analyzing power in
longitudinally polarized electron scattering from the proton at very low
$Q^{2}$ at an incident energy of $1.16$~GeV is in the final stages of
preparation for execution at Jefferson Laboratory (JLab). There exists an
unique opportunity to carry out the first ever precision measurement of the
weak charge of the proton, $Q^{p}_{W} = 1 - 4\sin^{2}\theta_{W}$, by making
use of the technical advances that have been made at JLab's world-leading
parity violating electron scattering program and by using the results of
earlier experiments to remove hadronic contributions. A $2200$ hour
measurement of the parity violating asymmetry in elastic electron-proton
scattering at $Q^{2}=0.03$~(GeV/c)$^{2}$ employing 180~$\mu$A of 85\% polarized
beam on a 0.35 m long liquid hydrogen target will determine the weak charge of
the proton with 4\% combined statistical and systematic errors. The Standard
Model makes a firm prediction of $Q^{p}_{W}$, based on the `running' of the
weak mixing angle $\sin^{2}\theta_W$ from the $Z^{o}$ pole down to lower
energies. Any significant deviation of $\sin^{2}\theta_W$ from its Standard
Model prediction at low $Q^{2}$ would constitute a signal of new physics.
In the absence of new physics, the envisaged experiment will provide a 0.3\%
determination of $\sin^{2}\theta_W$, making this a very competitive measurement
of the weak mixing angle. Complementary to the present experiment is a
measurement of the weak charge of the electron in parity violating M\o ller
scattering at $11$ GeV, currently under consideration, with the upgraded CEBAF
at JLab. The objective of that experiment would be a measurement of
$\sin^{2}\theta_W$ with a precision comparable to or better than any
individual measurement at the $Z^{o}$ pole.

\end{abstract}

\end{frontmatter}

\section{Introduction}
\label{intro}

Parity violating scattering of longitudinally polarized electrons from the
protons in a liquid hydrogen target allows the deduction of the weak analogues
of the conventional charge and magnetization distributions of the proton. In
turn one can extract the individual quark distributions from these form
factors. One should note the difference in the electromagnetic and weak
couplings to the quarks (see Table 1), pointing to the reversed sensitivities
of the proton and neutron. In going to lower and lower four-momentum transfers
the contributions due to the finite size of the proton become smaller and
smaller and one is able to measure then the weak charge of the proton, which
constitutes the sum of the weak charges of the two `up' quarks and the `down'
quark. However, the analyzing power becomes zero at zero four-momentum
transfer. A high precision measurement of the parity violating analyzing power
in elastic electron-proton scattering determines the value of
$\sin^{2}\theta_W$ and consequently the variation of $\sin^{2}\theta_W$ with
four-momentum transfer $Q^{2}$ or the `running' of $\sin^{2}\theta_W$. The
Standard Model makes a definitive prediction of the `running' of
$\sin^{2}\theta_W$ taking into account electroweak radiative corrections once
the value of $\sin^{2}\theta_W$ at the $Z^{o}$ pole has been reproduced. As
with the QED and QCD couplings, $\alpha(\mu^{2})$ and $\alpha_{s}(\mu^{2})$
(which exhibit screening and antiscreening, respectively), in going to higher
and higher four-momentum transfer, $\sin^{2}\theta_W$ is  an effective
parameter also varying with $\mu^{2} \approx Q^{2}$. In this case the
behaviour with $Q^{2}$ is more subtle since $\sin^{2}\theta_W$ is a function of
the electroweak couplings $g_{Vl}$ and $g_{Al}$: $(g_{Vl}/g_{Al}) = 1 -
4\sin^{2}\theta_W$. Any deviation of $\sin^{2}\theta_W$ from its Standard Model
predicted value points to new physics which needs to be incorporated through a
set of new diagrams. Measurements at the $Z^{o}$ pole have established the
value of the weak mixing angle $\sin^{2}\theta_W$ with great precision,
although it must be noted that the leptonic and semi-leptonic values of
$\sin^{2}\theta_W$ differ by $3\sigma$. The Standard Model `running' of
$\sin^{2}\theta_W$ has been calculated by Erler, Kurylov, and Ramsey-Musolf
\cite{erler} in the modified minimal subtraction scheme (see Fig.1). The
theoretical uncertainties $(\pm 0.00007)$  in the `running' of
$\sin^{2}\theta_W$ are represented by the width of the line. Hence the
interpretability is currently limited by the normalization of the curve at the
$Z^{o}$ pole, which is arguably as small as $\pm 0.00016$. Note the shift of
$+0.007$ at low $Q^{2}$ with respect to the $Z^{o}$ pole best fit of $0.23113
\pm 0.00015$. There have been reported several low energy measurements of the
value of $\sin^{2}\theta_W$. The first one is from an atomic parity violation
measurement in $^{137}$Cs \cite{bennett}, which agrees with the Standard Model
prediction within $1\sigma$ after many refinements detailing the atomic
structure of $^{137}$Cs were introduced. The second one is from a measurement
of parity violating M\o ller scattering \cite{anthony}, which also agrees with
the Standard Model prediction within approximately $1\sigma$. This is arguably
at present the better measurement in constraining extensions of the Standard
Model. The third one is from a measurement of neutrino and antineutrino
scattering from iron \cite{zeller} with a roughly $3\sigma$ deviation from the
Standard Model prediction. For this result there remain various uncertainties
in the theoretical corrections that need to be applied (among other two
identifiable effects of charge symmetry breaking in the quark distributions of
the nucleons \cite{londergan}). It is quite apparent that much higher precision
experiments are needed to search for possible extensions of the Standard
Model. One of these is a precision measurement of the weak charge of the
proton, $Q^{p}_{W} = 1 - 4\sin^{2}\theta_W$, currently being prepared for
execution in Hall C at JLab \cite{carlini}. The extraction of the value for
$\sin^{2}\theta_W$ is free of many-body theoretical uncertainties and has the
virtue of being able to reach much higher precision (note that, at a $Q^{2}$
value of $0.03$(GeV/c)$^{2}$, $1 - 4\sin^{2}\theta_W$ equals $0.07$). The
dominant hadronic effects that must be accounted for in extracting $Q^{p}_{W}$
from the measured analyzing power are contained in form factor contributions
which are sufficiently constrained from the current programs of parity
violating electron scattering (at MIT-Bates, JLab, and MAMI) without reliance
on theoretical nucleon structure calculations (note for instance the large
improvement in the knowledge of the neutral weak couplings to the valence
quarks that can be deduced from the current program of parity violating
electron scattering \cite{young94}). The Standard Model evolution of
$\sin^{2}\theta_W$ corresponds to a $10$ standard deviation effect in the
planned Qweak experiment at JLab. The Qweak experiment, the first ever
precision measurement of the weak charge of the proton and more precise than
the existing low energy measurements, is crucial in testing the Standard
Model. It is complementary to a parity violating M\o ller scattering
experiment, under consideration to be performed at $11$~GeV with an upgraded
CEBAF at JLab, with an envisaged precision in $\sin^{2}\theta_W$ equal to or
better than that from any individual measurement at the $Z^{o}$ pole. The
anticipated ($1\sigma$) uncertainties in both the Qweak experiment and a
future $11$~GeV M\o ller experiment are indicated in Fig.~1.  Needless to
remark: the electroweak radiative corrections to a pure leptonic measurement
are more contained. In the search for physics beyond the Standard Model,
precision measurements of the weak charge of the proton and of the weak charge
of the electron are rather complimentary.

\begin{table}
\begin{center}
\caption{ Electroweak charge phenomenology. The accidental
 suppression of the weak charge of the proton in the Standard Model gives
 it the better sensitivity to new physics.}
\begin{tabular}{lcr}
\hline
 & Electromagnetic Charge & Weak Charge \\
\hline
  $ q^{up}$           & +2/3 &  $1-(8/3)\sin^{2}\theta_W\approx 1/3 $ \\
  $ q^{down} $        & -1/3 &  $-1+(4/3)\sin^{2}\theta_W\approx -2/3 $ \\
  $ Q^{p}=2q^{up}+1q^{down}$ & +1 & $1-4\sin^{2}\theta_W=0.0716 $ \\
  $ Q^{n}=1q^{up}+2q^{down}$ &   0 &   -1   \\
\hline
\end{tabular}
\label{charge}
\end{center}
\end{table}
 
\begin{figure}
\begin{center}
\includegraphics*[width=10cm]{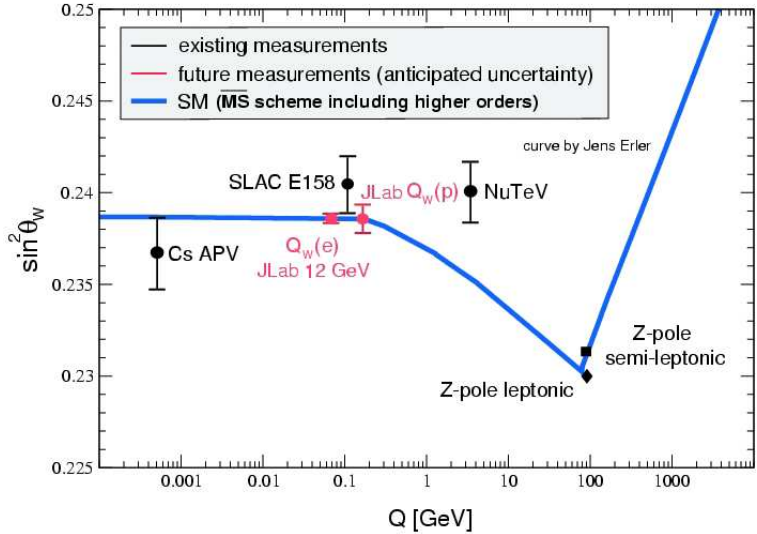}
\end{center}
\caption{Calculated `running' of the weak mixing angle in the Standard Model,
as defined in the modified minimal subtraction scheme \cite{erler}. The black
points with ($1\sigma$) error bars show the existing experimental values,
while the red points with error bars refer to the $4\%$ $Q^{p}_{W}$
measurement in preparation and a $2.5\%$ $11$~GeV M\o ller measurement under
consideration.
\label{zfig1}}
\end{figure}

\section{Overview of the Qweak Experiment}

The weak charge of the proton, $Q^{p}_{W} = 1 - 4\sin^{2}\theta_W$, will be
deduced from a precision measurement of the parity violating analyzing power
in elastic electron-proton scattering at a very low four momentum transfer
($Q^{2}$). The parity violating analyzing power is defined as:

\begin{displaymath}
A=(1/P)[\sigma^+ - \sigma^-]/[\sigma^+ + \sigma^-]
\end{displaymath}

where P is the polarization of the longitudinally polarized electron beam. It
was shown in \cite{musolf} that for forward angle scattering, where $\theta
\rightarrow 0$, the analyzing power can be written:

\begin{displaymath}
A=(1/P)\frac{-G_F}{4\pi\alpha\sqrt{2}}[Q^{2}Q^{p}_{W}+Q^{4}B(Q^{2})] 
\end{displaymath}

Here $G_F$ denotes the Fermi coupling constant and $\alpha$ is the fine
structure constant. One should note the dependence on $P$, which requires
precision polarimetry, and the dependence on the average value of $Q^{2}$
over the finite acceptance of the magnetic spectrometer based detector system
for scattered electrons, which requires the average value to be determined
through specific ancillary control measurements. The leading term in the
equation is the weak charge of the proton, $Q^{p}_{W} = 1 - 4\sin^{2}\theta_W$.
The quantity $B(Q^{2})$ represents the finite size nucleon structure and
contains the proton and neutron electromagnetic and weak form factors.  The
value of $B(Q^{2})$ can be determined experimentally by extrapolation from the
ongoing program of forward angle electron scattering parity violating
experiments at higher values of $Q^{2}$, already mentioned, or by specific
control measurements. The incident energy and the momentum transfer value
(mean scattering angle) followed from careful considerations of the figure of
merit. The optimum values are an incident energy of $1.165$~GeV and a momentum
transfer of $0.03$~(GeV/c)$^{2}$. One can then write for the longitudinal
analyzing power:
\begin{eqnarray}
A(0.03{\rm (GeV/c)}^{2}) &=&  A(Q^{p}_{W})+A(Had_V)+A(Had_A) \nonumber \\
                   &=&  -0.19~ppm-0.09~ppm -0.01~ppm   \nonumber
\end{eqnarray}
where the hadronic structure contributions are separated in vector and axial
vector components. Clearly, the total analyzing power is very small (-0.3
$ppm$) and one must arrive at an overall uncertainty of 2\% to meet the
precision objective of 0.3\% in $\sin^{2}\theta_W$. Consequently, high
statistics data are a prerequisite requiring high luminosity and high beam
polarization, and an integrating low-noise detector system of large
acceptance. As indicated above, the longitudinally beam polarization $P$ must
be precisely known as well as the hadronic structure contribution $B(Q^{2})$
to be subtracted from the measured analyzing power ($A(Had_V) + A(Had_A)$). A
better approach may be the fitting of all forward angle elastic scattering
electron-proton parity violation data as function of $Q^{2}$ which gives the
value of $Q^{p}_{W}$ at $Q^{2} = 0$. As in all parity violation experiments,
false analyzing power contributions result from helicity correlated changes in
the incident beam parameters, e.g., incident beam momentum and polarization,
intensity, position, direction, and width. The approach followed is to
minimize helicity correlated changes in the beam parameters, to design and
built a detector system as insensitive as possible to such changes (i.e., by
introducing cylindrical symmetry), and finally to measure the sensitivities of
the detector system and make corrections when necessary by measuring the
helicity correlated changes in the beam parameters during data taking.
Feedback loops will be introduced where and when absolutely necessary. The
Qweak experiment has set a goal of $6 \times 10^{-9}$ or less for helicity
correlated systematic error contributions to the analyzing power. All
backgrounds cause a dilution of the actual asymmetry and impose longer data
taking times if not yet accounted for. Backgrounds are minimized following
extensive simulations to define the proper collimator system and the
introduction of the appropriate shielding and optimization of the LH$_{2}$
target structure.

 The defining parameters of the Qweak experiment are given in Table~2. A
$2200$ hour measurement of the parity violating analyzing power in elastic
electron proton scattering at a momentum transfer of 0.03~(GeV/c)$^{2}$ with
180 $\mu$A of 85\% polarized beam incident on a 0.35 m long LH$_{2}$ target
will determine the weak charge of the proton with 4\% combined statistical and
systematic errors; this in turn will determine $\sin^{2}\theta_W$ at the 0.3\%
level at low $Q^{2}$. This approaches (by a factor of 2) the better individual
errors on $\sin^{2}\theta_W$ at the $Z^{o}$ pole in the SLD and LEP
experimental programs. A model independent analysis by Young et al.
\cite{young06} of published SAMPLE, PVA4, HAPPeX, and G0 data confirmed the
expected hadronic structure uncertainty entered in Table 3, which gives the
error budget for the Qweak experiment. The errors have been obtained through a
long process of extensive simulations and fully account for the effects of
Bremsstrahlung losses, including those inside the LH$_{2}$ target flask.

\begin{table}
\begin{center}
\caption{ Defining parameters of the Qweak experiment.}
\begin{tabular}{lr}
\hline
 Parameter & Value\\
\hline
 Incident beam energy & 1.165 GeV \\
 Beam polarization    & 85\% \\
 Beam current         & 180 $\mu$A \\
 Target thickness     & 0.35 m ($0.04X_{0}$) \\
 Data taking time     & 2200 hours \\
 Nominal scattering angle & $8.0^{o}$ \\
 Scattering angle acceptance & $\approx$ $\pm3.0^{o}$ \\
 Azimuthal acceptance & 53\% of $2\pi$ \\
 Solid angle          & $\Delta\Omega = 45 msr$ \\
 Average $Q^{2}$      & $0.028$~(GeV/c)$^{2}$ \\
 Average analyzing power & $-0.28 ppm$ \\
 Average experimental asymmetry & $-0.24 ppm$ \\
 Integrated cross section & $3.9$ $\mu$b \\
 Integrated rate (eight sectors) & $6.4$~GHz \\
 Statistical error on the asymmetry & 1.8\% \\
 Statistical error on $Q^{p}_{W}$ & 2.9\% \\
\hline
\end{tabular}
\label{param}
\end{center}
\end{table}

\begin{table}
\begin{center}     
\caption{ Total error estimate for the Qweak experiment. The contributions to
 both the parity violating analyzing power and the extracted $Q^{p}_{W}$ are
 given. The error magnification is due to the 39\% hadronic dilution.}
\begin{tabular}{lcr}
\hline
  Source of error & Contribution to & Contribution to \\
    & $\Delta$A/A & $\Delta Q^{p}_{W}/Q^{p}_{W}$ \\
\hline
  Statistical: \\    
  Counting statistics (2200 hours) & 1.8\% & 2.9\% \\
\hline
  Systematic: \\
  Beam polarimetry                 & 1.0\%     & 1.6\% \\
  Absolute $Q^{2}$                 & 0.5\%     & 1.1\% \\
  Helicity correlated beam parameter changes   & 0.5\% & 0.8\% \\
  Inelastic background uncertainty & 0.2\%     & 0.2\% \\
  Target window background         & $<$ 0.6\% & $<$ 0.8\% \\
  Hadronic structure uncertainties & ---       & 1.9\% - 2.4\% \\
  Radiative correction uncertainties in $Q^{p}_{W}$ & --- & $<$ 1\% \\
  Total systematic                 & 1.4\%     & 3.0\% \\
\hline
  TOTAL                            & 2.2\%     & 4.1\% - 4.3\% \\
\hline
\end{tabular}
\label{err}
\end{center}
\end{table} 

\section{Experiment Description}

The layout of the Qweak experiment is shown in Fig.~2. The main elements of the
Qweak experiment are a longitudinally polarized electron beam, a precision
collimator system, a resistive eight-fold symmetric toroidal magnetic
spectrometer, a set of eight detectors for the forward elastically scattered
electrons, and a set of luminosity monitors. The toroidal magnetic field will
focus the elastically scattered electrons onto the eight ersatz quartz
\v{C}erenkov detectors, each coupled on  either side to a photomultiplier tube
allowing read out in current mode for the high statistics Qweak data taking
and in counting mode for the ancillary $<Q^{2}>$ determination at greatly
reduced beam intensities (around 1 nA). Inelastically scattered electrons are
deflected out of the ersatz quartz detectors by the magnetic field of the
toroidal spectrometer. The defining parameters of the experiment are as given
in Table 2. The optimized kinematics correspond to an incident electron energy
of $1.165$~GeV and scattered electron polar angles of  $8.0 \pm \approx 3.0$
degrees. The azimuthal acceptance corresponds to 53\% of $2\pi$.

\begin{figure}
\begin{center}
\includegraphics*[width=10cm]{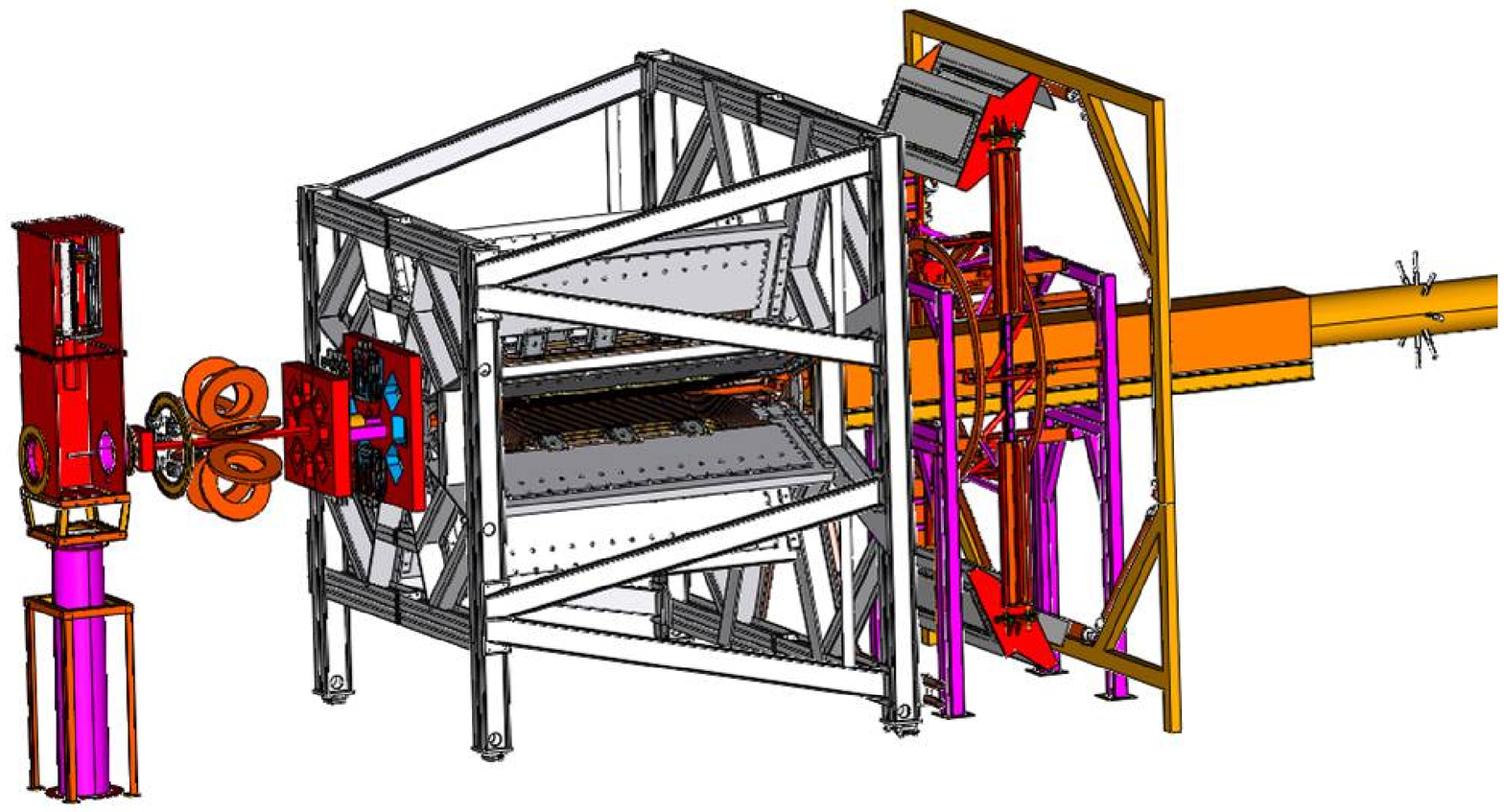}
\end{center}
\caption{Layout of the Qweak experiment. The beam is incident from the left
and scattered electrons exit the target and pass through the first collimator,
the region-1 GEM detectors, the two stage second precision collimator which
surrounds the region-2 drift chambers, the toroidal magnet, the shielding
wall, the region-3 drift chambers, the trigger scintillators, and finally the
ersatz quartz \v{C}erenkov detectors. The tracking system chambers and the
trigger scintillators, mounted on rotatable wheels, will be retracted outwards
during high current data  taking for the Qweak experiment proper. The
luminosity monitors, which will be used to monitor target density fluctuations
and to provide sensitive null asymmetry tests, are located downstream of the
main apparatus and are positioned very close to the through going beam.
\label{zfig2}}
\end{figure}

 The high current of the incident electron beam coupled to the length of the
LH$_{2}$ target that is required demand a cooling capacity of $2.5$~kW. This
presents the need for particular liquid helium cooling arrangements at JLab.
Target density fluctuations will be minimized by careful design of the
LH$_{2}$ target flask, by fast painting of the incident electron beam over the
LH$_{2}$ target (a raster of $4 \times 4$~mm in $5 \times 10^{-5}$~s), by a
fast circulation system (many liters per second), and by a spin flip frequency
of about $250$~Hz. The Monte Carlo simulations coupled to realistic tolerances
on the apparatus have resulted in a set of helicity correlated beam parameter
requirements which are given in Table 4.

\begin{table}
\begin{center}
\caption{ Helicity correlated beam parameter requirements for the Qweak
 experiment. The symbol $x_{0}$ refers to the DC beam position relative to the
 symmetry (neutral) axis of the apparatus; $\delta$x refers to the helicity
 correlated modulation of $x$; $r$ is the distance from the beam axis; and $D$
 is the beam diameter. These requirements should ensure that individual
 sources of systematic error produce false scattering asymmetries less than $6
 \times 10^{-9}$.} 
\begin{tabular}{lrrr}
\hline
 Helicity correlated & Error & Requirement & Requirement \\
 modulation & goes as & DC condition & helicity correlated limit \\
\hline
 Position & $x_{0}r^{2} \delta x$ & $x_{0}\leq3$~mm & $\delta x = 20$~nm \\
 Size     & $D_{0}^{3}  \delta D$ & $D_{0} = 4$~mm  & $\delta D \leq 0.7~\mu$m \\
 Direction & $\theta_{0}\delta\theta$ & $\theta_{0} =  60~\mu$rad & 
           $\delta\theta \leq 0.3~\mu$rad \\
 Energy  & $\delta$E & E = 1.165 GeV & $\delta$E/E $\leq 6 \times 10^{-9}$ \\
\hline
\end{tabular}
\label{beam}
\end{center}
\end{table}

 Downstream of the detection apparatus there are two sets of four luminosity
monitors each placed around the beam line at small angle, consisting of ersatz
quartz detectors coupled to radiation hard photodiodes with external
current-to-voltage converters. The small statistical error in the luminosity
detector signals allows corrections  for sensitivities to target density
fluctuations. The luminosity monitors will also provide a valuable asymmetry
null test since at their small angle the physics asymmetry has become
negligible small.

 The requirements on the main detector system are radiation hardness, low
sensitivity to different kinds of background, uniformity of response, and low
intrinsic noise. Following lengthy Geant-4 simulations, the choice has been
ersatz quartz (Spectrosil 2000, $n~=~1.47$) \v{C}erenkov bars of length
$2.0$~m,
of width $0.18$~m, and of thickness $0.0125$~m, for the detection of the
elastically scattered electrons. A shielding hut will protect the \v{C}erenkov
detectors from the significant ambient background present during data taking
at $180$ $\mu$A. The inelastic background contributing to the signal from the
\v{C}erenkov bars will be less than 1\%. Knowledge of the detector system
weighted $Q^{2}$ value will allow the inelastic background contribution to be
subtracted. A small quartz scanning detector is placed directly behind the
main detector bars and used as part of the acceptance mapping and linearity
testing at high and low incident electron beam currents.

 The Qweak tracking system consists of three sets of chambers. The upstream
region-1 chambers are Gas Electron Multiplier (GEM) chambers for fast response
and good position resolution. The region-2 chambers at the entrance to the
spectrometer, in between the defining collimators, are horizontal drift
chambers, while the region-3 chambers are vertical drift chambers just
upstream of the focal contour for the elastically scattered electrons, where
the \v{C}erenkov detectors are placed. The region-3 chambers will
momentum-analyze the particle trajectories. Finally, trigger scintillators are
installed between the region-3 chambers and the \v{C}erenkov bars in order to
provide a trigger to the electronics and a timing reference. The tracking
system will be able to determine the average $Q^{2}$ value to $\pm$0.5\% in
two opposing octants simultaneously. The three sets of chambers as well as the
trigger scintillators are mounted on three rotating wheel assemblies (shown in
Fig.~2) and can be retracted outwards (towards larger radii) during high
current data taking. Four sequential measurements with the tracking system are
required to map the entire detector system.

 The electron beam polarization needs to be measured with an accuracy less
than 1\%. This will be accomplished by upgrading the existing M\o ller
polarimeter in Hall C. The scheme adopted is for the high current polarized
electron beam to be deflected intermittently onto the polarized iron foil
containing the electrons with known polarization. In addition, a major effort
is underway to design, construct, and install a Compton polarimeter in Hall C,
which will allow continuous monitoring of the polarization of the electron
beam, but requires calibration against the M\o ller polarimeter. Both the
scattered electrons and back scattered photons will be detected.
 
\section{Conclusion}

 The Qweak experiment is a major undertaking at Jefferson Laboratory to measure
the weak charge of the proton with a precision that provides a significant
test of the Standard Model in the `running' of $\sin^{2}\theta_W$. Installation
of the Qweak instrumentation on the beam line in Hall C is slated to be
completed in 2009. Extensive simulations together with a rigorous program of
instrumentation design, construction, testing, and commissioning, and the
ongoing programs of measuring the hadronic form factor contributions, point to
the possibility of a 4\% measurement of the weak charge of the proton
translating into a 0.3\% measurement of $\sin^{2}\theta_W$.



\end{document}